\documentclass[twocolumn,preprintnumbers,amsmath,amssymb,superscriptaddress]{revtex4}

\usepackage{graphicx}
\usepackage{dcolumn}
\usepackage{bm}

\begin{document}


\title{Orbital ordering in cubic LaMnO$_{3}$ from first principles calculations}

\author{H. Zenia}
 \email{H.Zenia@Sheffield.ac.uk}
 \affiliation{Department of Physics and Astronomy, University of Sheffield, Sheffield, S3 7RH, UK }
 \affiliation{ Daresbury Laboratory, Daresbury, Warrington WA4 4AD, UK }
\author{G. A. Gehring}
 \affiliation{Department of Physics and Astronomy, University of Sheffield, Sheffield, S3 7RH, UK }
\author{W. M. Temmerman}
 \affiliation{ Daresbury Laboratory, Daresbury, Warrington WA4 4AD, UK } 

\begin{abstract}
We report on first principles Self-Interaction Corrected LSD (SIC-LSD)
calculations of electronic structure of LaMnO$_{3}$ in the cubic phase. We
found a strong tendency to localisation of the Mn $e_{g}$ electron and to
orbital ordering. We found the ground state to be orbitally ordered with a
staggered order of $x^{2}-z^{2}$ and $y^{2}-z^{2}$ orbits in one plane and
this order is repeated along the third direction. The difference in energy
with a solution consisting of the ordering of $3x^{2}-r^{2}$ and
$3y^{2}-r^{2}$ is, however, very small. 
The system is in the insulating A-type antiferromagnetic ordered state in both
cases. The presence of orbital ordering means breaking of the cubic
symmetry and without recourse to distortion. The latter may rather be the
result of the orbital ordering but the symmetry of this ordering is determined
by coupling to the lattice. The strong tendency to localisation of the $e_{g}$ 
electron in LaMnO$_{3}$ accounts for the survival of local distortions above
the structural phase transition temperature.

\end{abstract}


\maketitle

\section{\label{sec1} Introduction}
Orbital ordering in LaMnO$_{3}$ has always been associated with the Jahn-Teller 
instability of the system as a result of the degeneracy of the $e_{g}$
orbits. Because of the crystal field due to the oxygen octahedron the $t_{2g}$
orbits are known, from a local picture,  to lie lower in energy than the
$e_{g}$ ones. The $t_{2g}$ orbits are occupied each by one electron and because of
strong intraatomic Hund exchange interaction the spins of these electrons are
aligned parallel forming a $S=3/2$ spin which is sometimes treated as a
classical spin in model calculations. In  LaMnO$_{3}$ there is one more
electron that occupies an $e_{g}$ orbit. Because the two $e_g$ orbits are
degenerate the system is unstable towards a distortion which would lift the
degeneracy. The amount by which the system is distorted is then determined by
the competition between the gain in the electronic energy and the increase in
the elastic energy of the lattice due to distortion. LaMnO$_{3}$ is found in
the distorted phase below 780K. Orbital ordering in LaMnO$_{3}$ has been
observed by resonant X-ray scattering on  the Mn-K edge. It was found that
this ordering decreases above the N\'eel  temperature 140K and disappears
above T=780K concomitant with a structural phase transition\cite{murakami}.

In 3d transition metal compounds with orbital degeneracy two scenarios
are invoked to explain orbital ordering. On the one hand there is superexchange
interaction between orbitals on different sites. This involves virtual
transfer of electrons and strong on-site electron-electron interaction. On the
other hand cooperative JT distortions or electron-lattice interaction leads to
splitting of the degenerate orbits and thus to orbital ordering. Although the
ground state ordering of LaMnO$_{3}$ can be explained by both mechanisms it is
not easy to say which is the dominant contribution. This question may sound of
little importance so far as LaMnO$_{3}$ is concerned but it is important to
know the answer because whichever is the dominating mechanism will remain more
or less active once the system is doped. In which case the two
mechanisms, \textit{i.e.}, electron-lattice interactions or electron-electron correlations may
lead to different physics for the doped systems.

There have been attempts, by using model calculation, at explaining how
orbital ordering can occur if one assumed an antiferromagnetic spin ordering
\cite{efremov}. However, as mentioned above, temperatures at which
the orbital ordering sets in are much higher than the N\'eel temperature of
A-AF spin ordering. Orbital ordering can not, therefore, be attributed to spin
ordering. In previous LSD \cite{solovyev,pickett,satpathy} and model HF
\cite{mizokawa95,mizokawa96} calculations it is found that inclusion of
distortions is necessary to recover the correct A-AF and insulating character
of LaMnO$_{3}$ in the ground state. The cubic system was found to be both
metallic and ferromagnetic in the LSD calculations. By using Self-Interaction
Correction (SIC) to the LSD we can allow for the $t_{2g}$ orbitals to localise
and  form low lying semi-core manifold well below the $e_{g}$ orbits
\cite{zenia}. We then can  compare total energies for different scenarios
corresponding to localising a  particular $e_{g}$ orbit. By doing so one
breaks the cubic symmetry but this is allowed if the resulting  ground state
is lower in energy.  As a result of the orbital ordering the system will
distort in order to reduce the electrostatic energy due to the interaction of
the oxygen electronic clouds with the lobes of the $e_g$ orbit, that is
directed to them, on neighbouring Mn ions. 

\begin{table*}[ht]
\caption{\label{tab1} Total energies in mRy per formula unit and magnetic moments in
  $\mu_{B}$ of  cubic LaMnO$_{3}$ in the FM, A-AFM  and G-AFM
  magnetic orderings with several orbital ordering scenarios. Where one orbit
  only is specified the orbital ordering is ferro and for two orbits e.g.
  $3x^{2}-r^{2}$/$3y^{2}-r^{2}$  the ordering is of a C-type with the ordering
  vector ${\bf q} = \frac{\pi}{a}(1,1,0)$. The
  energies are given as differences with respect to the energy of the
  solution corresponding to the experimentally known structure of the
  distorted LaMnO$_{3}$.}
\begin{ruledtabular}
\begin{tabular}{l|cccccccc}
 Configuration &&$lsd$        &$t_{2g}$  &$3z^{2}-r^{2}$ &$x^{2}-y^{2}$
                                 &$3x^{2}-r^{2}$/$3y^{2}-r^{2}$ &$x^{2}-z^{2}$/$y^{2}-z^{2}$ &$3x^{2}-r^{2}$/$3z^{2}-r^{2}$ \\
\colrule 
 &   FM            &140.3      &21.4      &8.1           &11.7
                                 &0.5                         &-0.5                         &6.3\\
Energy&    A-AFM         &152.0      &30.7      &7.2           &11.4 
                                 &0.0                         &-0.6                         &4.9\\
 &   G-AFM         &160.9      &45.2      &9.4           &9.5
                                 &7.5                         &7.7                          &8.8 \\
\colrule
\colrule
    &    FM            &2.89       &3.07      &3.72          &3.70    
                                 &3.70                        &3.71                         &3.68 (3.72)\\ 
Mn mom.&   A-AFM         &2.81       &3.14      &3.62          &3.69
                                 &3.70                        &3.67                         &3.67 (3.64) \\     
    &   G-AFM         &3.10       &3.41      &3.60          &3.60
                                 &3.60                        &3.61                         &3.63 (3.57)\\
\end{tabular}
\end{ruledtabular}
\end{table*}

The basis states are written: $|x\rangle=x^{2}-y^{2}$ and $|z\rangle=\frac{1}{\sqrt{3}}(2z^{2}-x^{2}-y^{2})$.
A composite state can be written as:
$|\theta\rangle=\cos\frac{\theta}{2}|z\rangle+\sin\frac{\theta}{2}|x\rangle$ \cite{sikora}.
Then the orbital state $|z\rangle$ corresponds to $\theta=0$ and the state
$|x\rangle$ to $\theta=\pi$. The orbital ordering of LaMnO$_{3}$ consists of
an antiferro ordering of two orbits, viz.,
$|\pm\theta\rangle=\cos\frac{\theta}{2}|z\rangle\pm\sin\frac{\theta}{2}|x\rangle$
in a plane while the same order is repeated along the 
third direction. Until recently it was assumed that $\theta=2\pi/3$. But recent ESR 
\cite{deisenhofer} and neutron diffraction \cite{rodriguez-carvajal} measurements have etimated $\theta$ to be 
$92^{o}$ and $106^{o}$ respectively. Phenomenological superexchange calculations for the ground state ordering have
also given $\theta_{opt}\sim83^{o}$, ``significantly different from $2\pi/3$'' \cite{sikora}. Our current
calculations are however limited to the cases of ferro order of $\theta=$$0$
and $\pi$ and antiferro order of $\pm\pi/3$ and $\pm2\pi/3$.

\section{\label{sec2} Calculation details}
The calculations are performed in the SIC-LSD approximation implemented within
the LMTO-ASA method \cite{perdew,temmerman}. The SIC corrects for the
spurious interaction of an electron with itself inherent to the LDA
approximation to the exchange correlation functional of the DFT. It is known
however that this energy is important only when the electron is localised
whereas it vanishes for delocalised electrons. This method is used to determine
whether it is favourable for an electron to localise or to be
itinerant. This is done by comparing the total energies of the system in the
presence of the two scenarios. The lattice parameter used in the present
calculation, $a_{0}=7.434$a.u., is the one which gives the experimental volume of the
real distorted LaMnO$_{3}$ system. We have used a minimal basis set consisting
of $6s$, $5p$, $5d$ and $4f$ for La, $4s$, $4p$ and $3d$ for Mn and $2s$,
$2p$, and $3d$ for O. Mn $4p$ and O $3d$ were downfolded. For the atomic
sphere radii we used 4.01, 2.49 and 1.83 a.u for La, Mn and O
respectively. In order to look at different orientations of the two orthogonal
$e_{g}$ orbitals we used rotations of the local axes on the Mn sites. We checked the
accuracy of these rotations by comparing the total energies of three
configurations: all $3z^{2}-r^{2}$, all $3x^{2}-r^{2}$ and all $3y^{2}-r^{2}$
localised in both FM and G-AFM cases because these magnetic orderings preserve
the cubic symmetry and hence the energy should not be dependent on which orbit is
localised so long as it is the same one on all the Mn sites. The
energy differences found in this way were always less than 1mRy per
formula unit.

The calculations were done for a four-formula unit cell. The notations of the
orbital ordering scenarios are as follows: $lsd$: LDA calculation with no
self-interaction correction; $t_{2g}$: SIC applied to the $t_{2g}$ orbits only
on all the Mn sites, and in all the other cases one $e_{g}$ orbit is localised
on top of the $t_{2g}$ ones. The remaining scenarios correspond to
localising either the same or different orbits in the $ab$ plane while
preserving the same ordering on the second plane along $c$. Thus we have
either ferro or C-type antiferro orbital ordering.

\section{\label{sec3} Results and discussion}

From the total energies of Table \ref{tab1} we see that the ground state
corresponds to an orbitally ordered solution forming a C-type antiferro-orbital arrangement
of the $x^{2}-z^{2}$  and $y^{2}-z^{2}$ in the $ab$ plane with the same
ordering repeated along the $c-$axis. The corresponding magnetic ordering is of
A-type AFM as found in the distorted system. This solution is however almost
degenerate with the solution with an ordering of $3x^{2}-r^{2}$  and
$3y^{2}-r^{2}$. The energy difference between the two solutions, 0.6mRy/f.u., is within the
accuracy of the calculation method(LMTO-ASA). It is then most likely that the
true ground state of the cubic system is made up of a combination of both
solutions. Interactions with the neighbouring oxygens are certainly
different for the two orderings and relaxation of the oxygen positions in the
real system may favour one of the solutions or a linear combination of
them.

\begin{table*}[ht]
\caption{\label{tab2} Magnetic exchange constants in meV obtained from the
  total energies in Table \ref{tab1}. $J_{1}$ and $J_{2}$ are Heisenberg
  in-plane and inter-plane exchange integrals respectively.}
\begin{ruledtabular}
\begin{tabular}{l|ccccc}
OO scenario &$3z^{2}-r^{2}$ &$x^{2}-y^{2}$
          &$3x^{2}-r^{2}$/$3y^{2}-r^{2}$ &$x^{2}-z^{2}$/$y^{2}-z^{2}$ &$3x^{2}-r^{2}$/$3z^{2}-r^{2}$ \\  
\colrule
$8J_{1}S$   &14.96  &-12.93 &51.02  &56.46  &26.53 \\
\colrule
$4J_{2}S$   &-6.12  &-2.04  &-3.40  &-0.68  &-9.52 \\
\end{tabular}
\end{ruledtabular}
\end{table*}

We have considered three types of spin order.  
Ferromagetism and A type antiferromagnetism where the spins are parallel in the x-y planes 
and the planes are stacked antiparallel up the z axis and 
G type antiferromagnetism where each spin is antiparallel to all its neighbours.
The difference in energy between 
the FM and A-AFM magnetic orderings in the two cases is also very small which
is consistent with the fact that inter-plane AF exchange is much smaller than
in-plane FM exchange in agreement with experiments. Experimental exchange integrals
are obtained from fitting neutron scattering results (spin wave
dispersion) to a simple Heisenberg Hamiltonian with two exchange integrals
acting between nearest neighbours. We calculated the exchange constants using the
convention of Ref \cite{hirota}: $E_{F}=(-4J_{1}-2J_{2})S^{2}$,
$E_{A-AF}=(-4J_{1}+2J_{2})S^{2}$ and $E_{G-AF}=(4J_{1}+2J_{2})S^{2}$ for the
energies of the FM, A-AFM and G-AFM respectively. We assumed the value of
$S=2$ for the magnetic moment on Mn ions for all the orderings. The results are given, in Table 
\ref{tab2}, for different orbital ordering (OO) scenarios of the $e_{g}$
orbits as given in Table \ref{tab1}. Experimentally the two exchange integrals
are found to be $8J_{1}S=13.36\pm0.18$meV and $4J_{2}S=-4.84\pm0.22$meV
for the in-plane and inter-plane coupling respectively
\cite{hirota,moussa}. 
We see then that our
calculation overestimates the tendency to in-plane ferromagnetism whereas the
interplane exchange is marginally underestimated. However it was found in LDA calculations \cite{solovyev}
that the first neighbour exchange integrals depend dramatically on lattice
distortions. This  might explain why our exchange constants calculated
for the cubic lattice are quantitatively different from the experimental ones
which were determined for the distorted lattice. Our results are however in disagreement with recent model 
calculations of Sikora and Ole\'{s} \cite{sikora} who found that for an ordering of ``$\theta=2\pi/3$, often assumed
for LaMnO$_{3}$,'' the exchange constants ``are never close to the experiment''. Their calculated constants are both
ferromagnetic which contradicts the experimental fact that LaMnO$_{3}$ is an A-type antiferromagnet. Hence their 
argument that $\theta$ should in fact be different from the assumed $2\pi/3$.

The widely used Goodenough-Kanamori (G-K) rules \cite{goodenough63,khomskii01} 
give an indication of which exchange interactions 
should be positive (ferromagnetic) and which negative (antiferromagnetic) 
depending on the state of ionisation of the two ions, 
the occupied orbitals and the angle subtended at the bridging ion.  
They are valid only for insulating states and were worked out using perturbation theory to give a 
general guide to the interactions although deviations are known to occur.\cite{meskine01}  
It is useful to compare our results for this specific material with the predictions of the G-K rules 
because the results may be used in future to assess the reliability of the rules.  
We note that in the case where we have ferromagnetism and ferromagnetically aligned orbits 
then our results predict a metallic ground state and so in these cases the rules are not applicable. 

In  LaMnO$_{3}$ the Mn ions are all in the same oxidation state and the Mn ions and the bridging oxygen 
lie along a straight line in the cubic unit cell; the bridging angle is $\pi$.  
Thus the only variable that is relevant to the G-K rules is the orbital order.  
The rules state that if nearest neighbour sites are occupied by the same orbit the interaction is 
negative, antiferromagnetic.  The size of the effect depends directly on the overlap 
of the orbits e.g. if there are two orbits $3x^{2}-r^{2}$ (which have large lobes in the x direction) 
separated by a lattice vector directed along x it would be larger than if, for example, 
the two orbits were $3y^{2}-r^{2}$ but still separated by a lattice vector directed along x.  
This would fit nicely with the result in Table \ref{tab2} that the value of $J_{2}$ (exchange up the z direction)  
is negative for OO $3x^{2}-r^{2}$/$3y^{2}-r^{2}$ and also for OO $3x^{2}-r^{2}$/$3z^{2}-r^{2}$ 
but significantly larger in the latter case where there are $3z^{2}-r^{2}$ orbits arranged 
in columns up the z axis.  The calculation of $J_{1}$ is more complicated\cite{meskine01} 
because the orbits are partially occupied but is ferromagnetic 
for OO $3x^{2}-r^{2}$/$3y^{2}-r^{2}$.\cite{khomskii97}  
The overlaps would be smaller in the x-y plane for the case OO $3x^{2}-r^{2}$/$3z^{2}-r^{2}$ 
than for OO $3x^{2}-r^{2}$/$3y^{2}-r^{2}$ so application of the G-K rules 
would predict a larger value of $J_{1}$ (exchange in the x-y plane) 
in the former case in agreement with first principles results.  
The signs of $J_{1}$ and $J_{2}$ in Table \ref{tab2} do agree with the G-K rules.
There is one detail in which the first principles results do deviate 
from the G-K rules and that is in the case antiferromagnetic ordering with 
the ferromagnetic orbital order F $3z^{2}-r^{2}$. In this case 
since all the orbits are the same all the nearest neighbour interactions should be antiferromagnetic 
which would mean that the G-AF state should be more favourable than the A-AF state 
whereas the opposite order is seen in Table \ref{tab1} and 
the value for $J_{1}$ in Table \ref{tab2} should be negative for this orbit.
The order is correct for the other ferromagnetic orbital order, F $x^{2}-y^{2}$.
Thus we see the predictions for the signs of $J_{1}$ and $J_{2}$ from the first principles calculation 
and the G-K rules agree in all cases (except the one mentioned for F $3z^{2}-r^{2}$ orbit above) 
and the magnitudes agree.  
In one case we see that there is a disagreement on the ordering of unfavourable states. Model perturbation
calculations of the exchange constants also disagree with the G-K rules: As mentioned earlier, Sikora and Ole\'{s} 
\cite{sikora} have found that for the case of $\theta=2\pi/3$ the constants are small and both ferromagnetic, 
whereas G-K rules predict that J$_{1}$ is strongly ferromagnetic while $J_{2}$ is antiferromagnetic.

In the ferromagnetic
case the total moment is $4\mu_{B}$ which is the value one expects
from having four $d$ electrons. 
This is the case also because the FM
solution is either half-metallic (OO scenarios $3z^{2}-r^{2}$ and $x^{2}-y^{2}$ of Table
\ref{tab1}) or insulating
(OO scenarios $3x^{2}-r^{2}$/$3y^{2}-r^{2}$, $x^{2}-z^{2}$/$y^{2}-z^{2}$ and 
$3x^{2}-r^{2}$/$3z^{2}-r^{2}$ of Table \ref{tab1}). 
The magnetic moment that is on the Mn
ion can be less than this because of hybridiation . The magnetic moment on the Mn
ion when one $e_{g}$ orbit is localised is about 3.70$\mu_B$ in both FM and
A-AFM solutions and of 3.60$\mu_B$ in the G-AFM case. Because of hybridisation
with the oxygen part of the polarisation is sitting on the oxygen ion. 

\begin{figure}[h]
\includegraphics[trim = 0mm 0mm 0mm -20mm, scale=0.35]{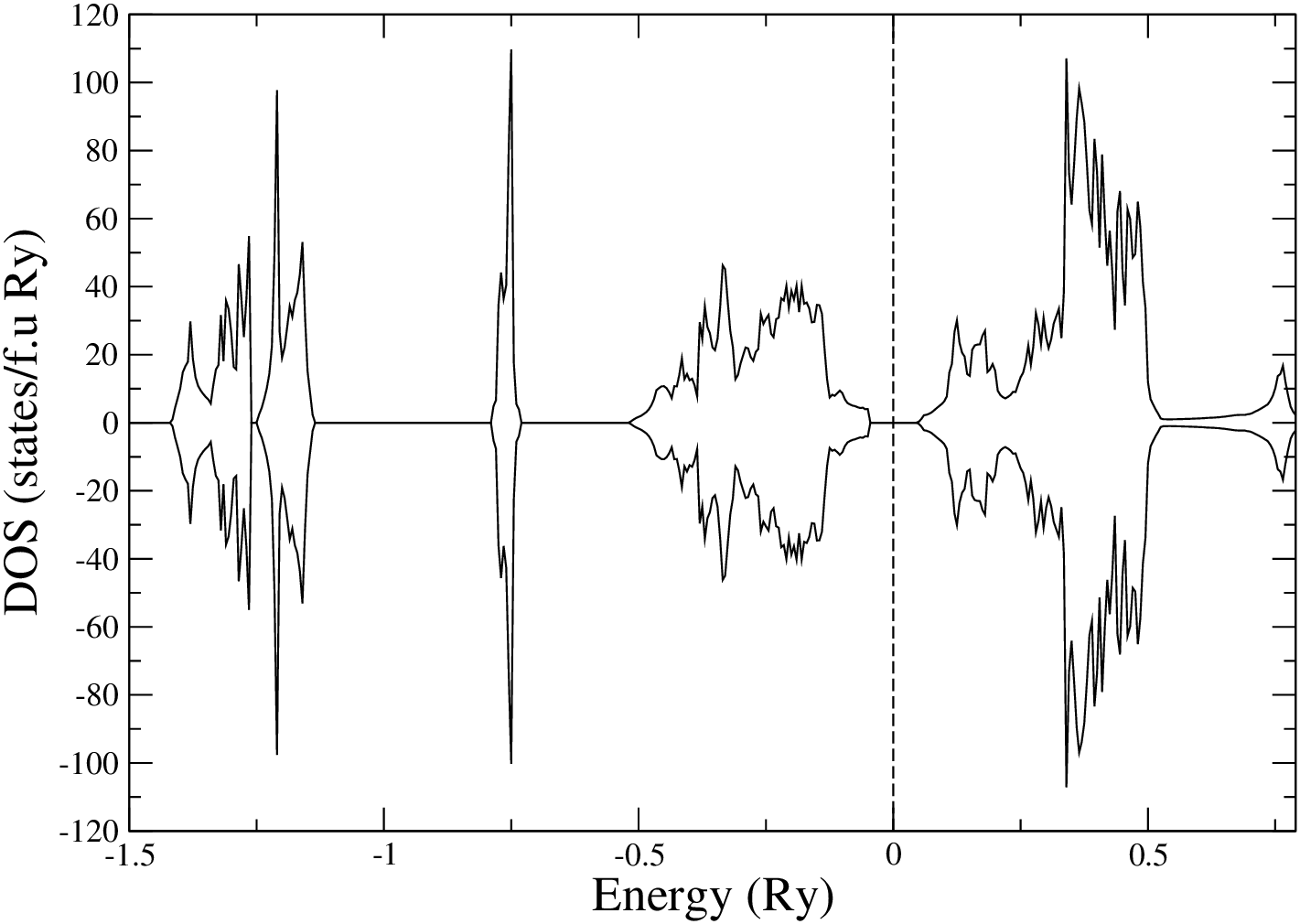}
\caption{\label{fig1} Total DOS in A-AFM magnetic and $x^{2}-z^{2}$/$y^{2}-z^{2}$
  orbital orderings.}
\end{figure}

\begin{figure}
\includegraphics[trim = 0mm 0mm 0mm -20mm, scale=0.35]{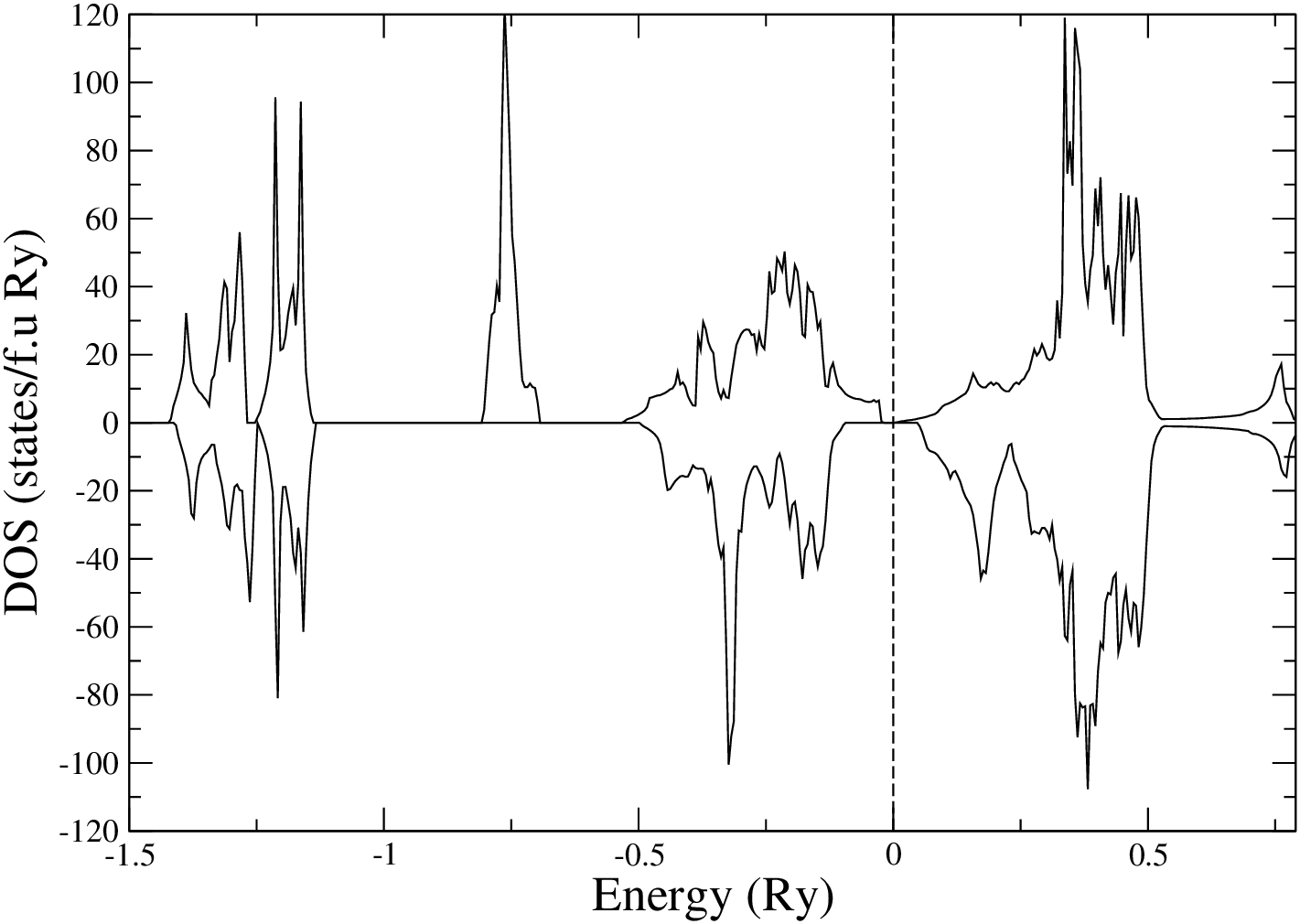}
\caption{\label{fig2} Total DOS in FM magnetic and $x^{2}-z^{2}$/$y^{2}-z^{2}$
  orbital orderings.}
\end{figure}

\begin{figure}[h!]
\includegraphics[trim = 0mm 0mm 0mm -20mm, scale=0.35]{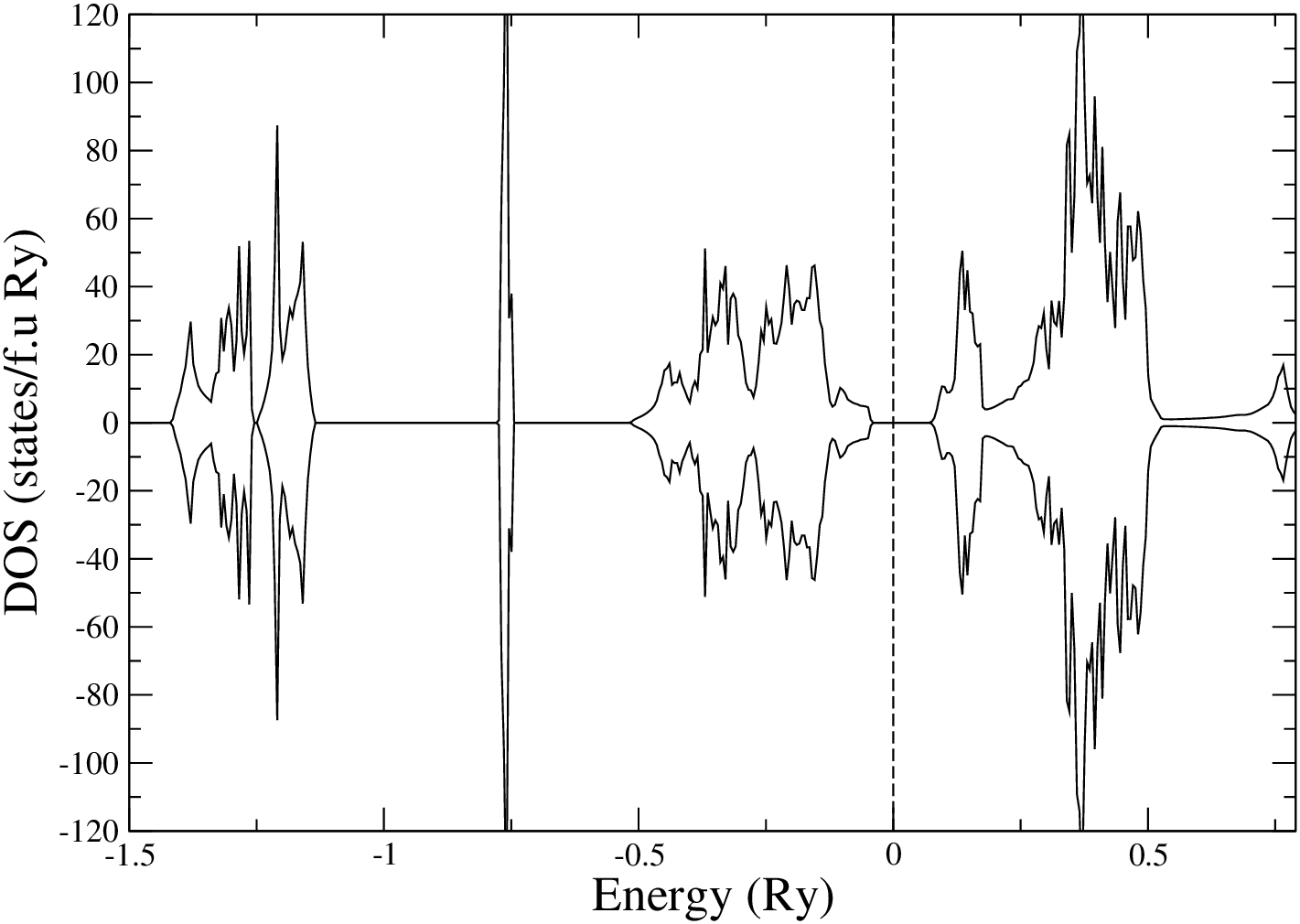}
\caption{\label{fig3} Total DOS in G-AFM and $x^{2}-z^{2}$/$y^{2}-z^{2}$
  orbital orderings.}
\end{figure}

\begin{figure}[h!]
\includegraphics[trim = 0mm 0mm 0mm -20mm, scale=0.35]{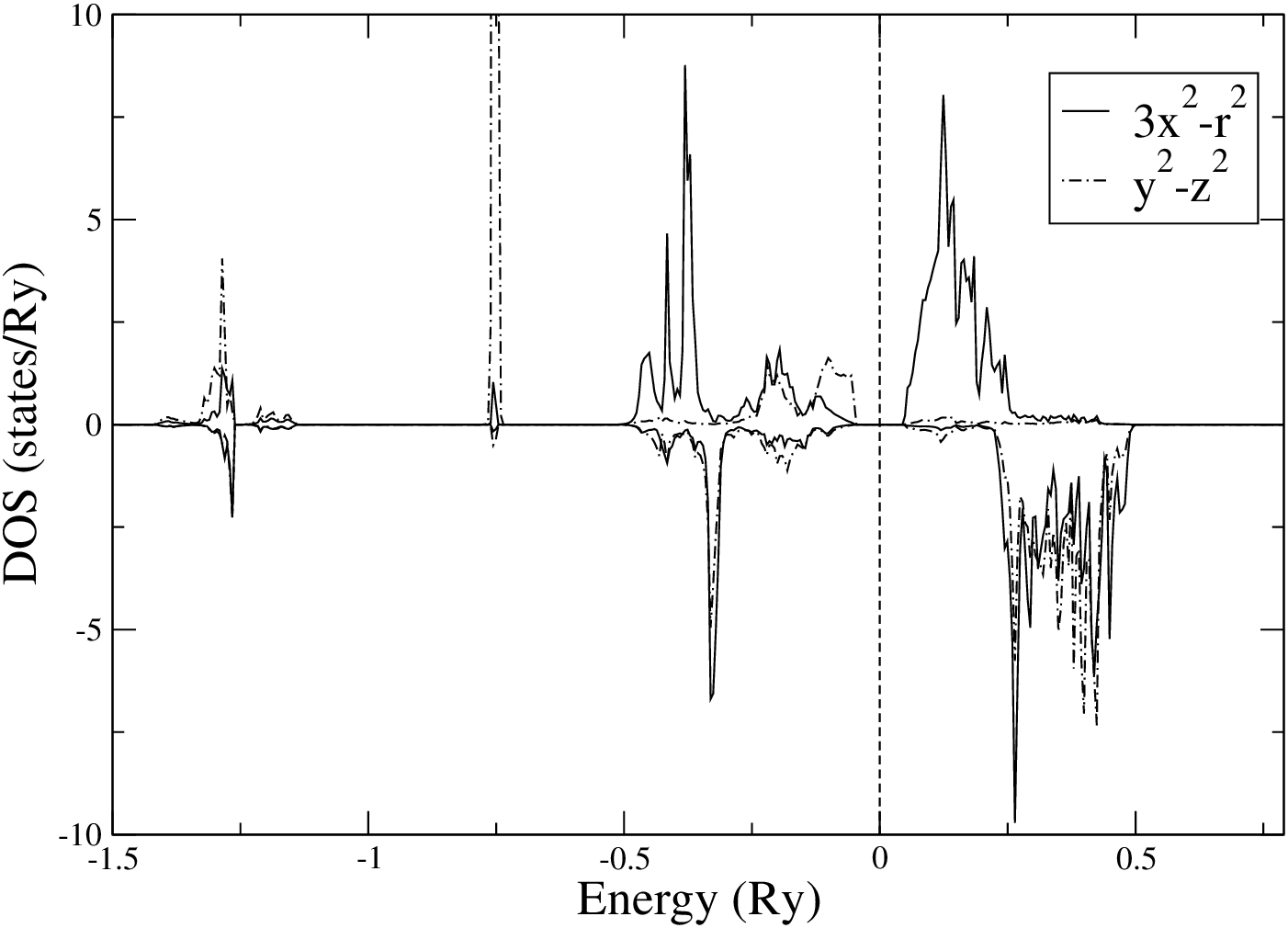}
\caption{\label{fig4} Mn $e_{g}$-projected DOS in the ground state A-AFM in the
  $x^{2}-z^{2}$/$y^{2}-z^{2}$ orbital ordering on the Mn site with the self interaction correction applied
to the $y^{2}-z^{2}$ orbital.}
\end{figure}

The system is  insulating in both orbital ordering scenarios independently of the magnetic 
ordering. Inspection of the total density of states (DOS) in the lowest energy
$x^{2}-z^{2}$/$y^{2}-z^{2}$ ordering scenario presented in
Figs. \ref{fig1}(A-AFM), \ref{fig2}(FM) and \ref{fig3}(G-AGM) reveals the
presence of a gap which is larger as more nearest neighbour spins become
antiferromagnetic (See also Table \ref{tab3}). Its
calculated value in the $3x^{2}-r^{2}$/$3y^{2}-r^{2}$ orbital and AFM magnetic
orderings 
is in very good agreement with the experimental optical gap \cite{arima} as can be seen in
Table \ref{tab3}. The peak at about -0.75Ry in the total DOS corresponds to the localised 3$t_{2g}$
and one $e_{g}$ orbits. The latter are shown in Fig. \ref{fig4} where
we can see the following features in the majority spin channel: the peak at -0.75Ry representing the
localised $y^{2}-z^{2}$ states 
and the $3x^{2}-r^{2}$ states split into occupied states which hybridize strongly with the oxygen $2p$ 
states and unoccupied $3x^{2}-r^{2}$ states. 
One can also notice by looking at the minority
$e_{g}$ states that both orbits are degenerate because these are not
corrected for by the SIC and hence are solutions of the LSD
potential which are orthogonal to the SIC states. 
In the LSD calculation the $t_{2g}$ and $e_{g}$ states lie near the Fermi
level with the $t_{2g}$ states somewhat more localised than the
$e_{g}$ ones. 
However the LSD does not describe their localisation
accurately. 
In the SIC they are pushed well below the valence band, composed mostly of Oxygen
$2p$ states. It is however known that the position of the
SI-corrected levels does not correspond to what would be seen in
experiment. Relaxation effects need to be considered if one wanted to
get spectra from SIC single particle energies \cite{temmerman_pr}. Centred around -1.25Ry
are the oxygen $2s$ and La $5p$ semi-core levels.

\begin{table}[h]
\caption{\label{tab3} Energy band gaps in eV.}
\begin{tabular}{|c|c|c|c|}
\colrule 
\colrule 
 Configuration   &$3x^{2}-r^{2}$/$3y^{2}-r^{2}$ &$x^{2}-z^{2}$/$y^{2}-z^{2}$   &Exp \\
\colrule 
 FM             &0.54                         &0.27                          &    \\
\colrule
 A-AFM          &1.09                         &1.29
 &1.1\footnote{Ref. \cite{arima}} \\
\colrule
 G-AFM          &1.50                         &1.56                          &\\
\colrule 
\colrule 
\end{tabular}
\end{table}

The total energy of the solution where only $t_{2g}$ orbits are localised and the $e_{g}$ electron 
is delocalised lies much higher than the most unfavourable orbital ordering solution which confirms 
that there is strong tendency to the localisation of the $e_{g}$ electron in LaMnO$_{3}$
even in the cubic phase. The energy scale of the
localisation/delocalisation of the  $e_{g}$ electron is indeed at
least twice as big as the energy corresponding to ordering the
orbits. This is qualitatively in agreement with the experimental
observation that even above the critical temperature of the orbital
ordering local distortions remain. Local distortions are an indication
that there is localisation. Once these $e_{g}$ electrons are
localised they induce local distortions through the interactions with the
surrounding oxygens and these distortions order simultaneously with the
orbits when the temperature is lowered. Although we can
not with the current method simulate real paramagnetism as being a collection
of disordered local moments without long range ordering we can speculate however, since
the orbital ordering is so strong and independent of the spin ordering, that
orbital ordering occurs in the paramagnetic state too. It is this orbital
ordering which drives magnetic ordering and not the other way round. In a
model calculation of paramagnetic LaMnO$_{3}$ and KCuF$_{3}$ based on an LDA+U
electronic structure Medvedeva \textit{et al.}  \cite{medvedeva} concluded
that distortions were not needed to stabilise the orbitally ordered phase in
both compounds and that this ordering is of purely electronic origin. Their
calculations for cubic LaMnO$_{3}$ have found that in the PM phase the orbits
order but they are not pure local $3z^{2}-r^{2}$ and $x^{2}-y^{2}$. They found
that the local $3z^{2}-r^{2}$ has an occupancy of 0.81 and the local
$x^{2}-y^{2}$ has an occupancy of 0.21. This is consistent with our present
calculations in that the calculated ground state is nearly
degenerate. 
In earlier LDA+U calculations \cite{liechtenstein} on KCuF$_{3}$
it was found that within LDA there was no instability of the system against
distortion while in LDA+U the energy has a minimum for a finite distortion of
the lattice. It was concluded then that electron-phonon and exchange only are
not enough to drive the collective distortion. A similar view was supported
also by model calculations \cite{mostovoy,okamoto} where both
electron-electron and electron-lattice interaction are taken into account. In
our present calculation the competition is rather in terms of
localisation/delocalisation of the $e_{g}$ orbits by electronic interactions
alone. And we found indeed that these are enough to first localise the orbits
(larger energy scale) and then to order them in an anti-ferromagnetic
way(smaller energy scale). Based on these results and those mentioned earlier
we speculate that the distortions are a consequence of the displacement of
oxygen ions to accommodate the electrostatic interactions resulting from the
orbital ordering but these are crucial in selecting the ground state ordering
out of the two nearly degenerate solutions we found for the cubic case.

Earlier SIC-LSD calculations by Tyer \textit{et al.} \cite{Rik}
have described correctly the physics of the
distorted LaMnO$_{3}$. Then Banach and Temmerman \cite{Banach} studied
the cubic phase but using a unit cell of two formula units only. This
limited the study to the first two rows and first four columns of
Table \ref{tab1}. Hence they found that the lowest energy solution is
the A-AFM with $3z^{2}-r^{2}$ orbital ordering. Upon decreasing the
lattice parameter they found a crossover to the FM with $t_{2g}$
orbitals SI-corrected only which means suppression of orbital
ordering. We reconsidered this case below with our present
bigger cell.

Loa \textit{et al.} \cite{loa} studied structural and electronic properties of
LaMnO$_{3}$ under pressure and found that the system is still insulating even
at higher pressure than the critical one at which the structural transition takes place.
There was no indication of the magnetic state of the system but the
experiments were carried out at room temperature which is well above the
ordering temperature at least of the distorted LaMnO$_{3}$. We found both FM
and A-AFM solutions to be insulating in both $3x^{2}-r^{2}$/$3y^{2}-r^{2}$ and
$x^{2}-z^{2}$/$y^{2}-z^{2}$ orbital ordered states. Whereas the system is
metallic when only the $t_{2g}$ electrons are localised. The fact that the
system was found to be insulating after suppression of the JT distortion is
indicative of the presence of orbital ordering with or without spin
ordering. Use of local probe such as
EXAFS or PDF(pair distribution function) would be of great help though to
settle the question of whether pressure really quenches distortions at the
local level.

Another way of suppressing the distortions is by increasing temperature as
done by S\'anchez \textit{et al.} \cite{sanchez} who studied the structural
changes of LaMnO$_{3}$ with temperature by using XANES and EXAFS measurements.
Probing the local environment of the Mn ions they found no abrupt change in
the signal upon crossing the structural transition temperature T$_{JT}$. They
described the structural phase transition as ordering of the local distortions
that are thermally disordered above T$_{JT}$ resulting in a cubic lattice
on average. This picture is quite different from the high pressure one
although in both cases the distortions are apparently suppressed. In the high
temperature regime orbital ordering can still be present but the long range
ordering is suppressed by thermal fluctuations. Consistent with our
calculation that the localisation/delocalisation energy is of a larger scale
than orbital ordering, i.e, the $e_{g}$ electrons tend to localise
strongly. As a consequence  the lattice is distorted locally but since the
energy scale of ordering the orbits/distortions is lower they are disordered
by thermal fluctuations at high temperature.

\begin{figure} 
\includegraphics[trim = 0mm 0mm 0mm -20mm, scale=0.35]{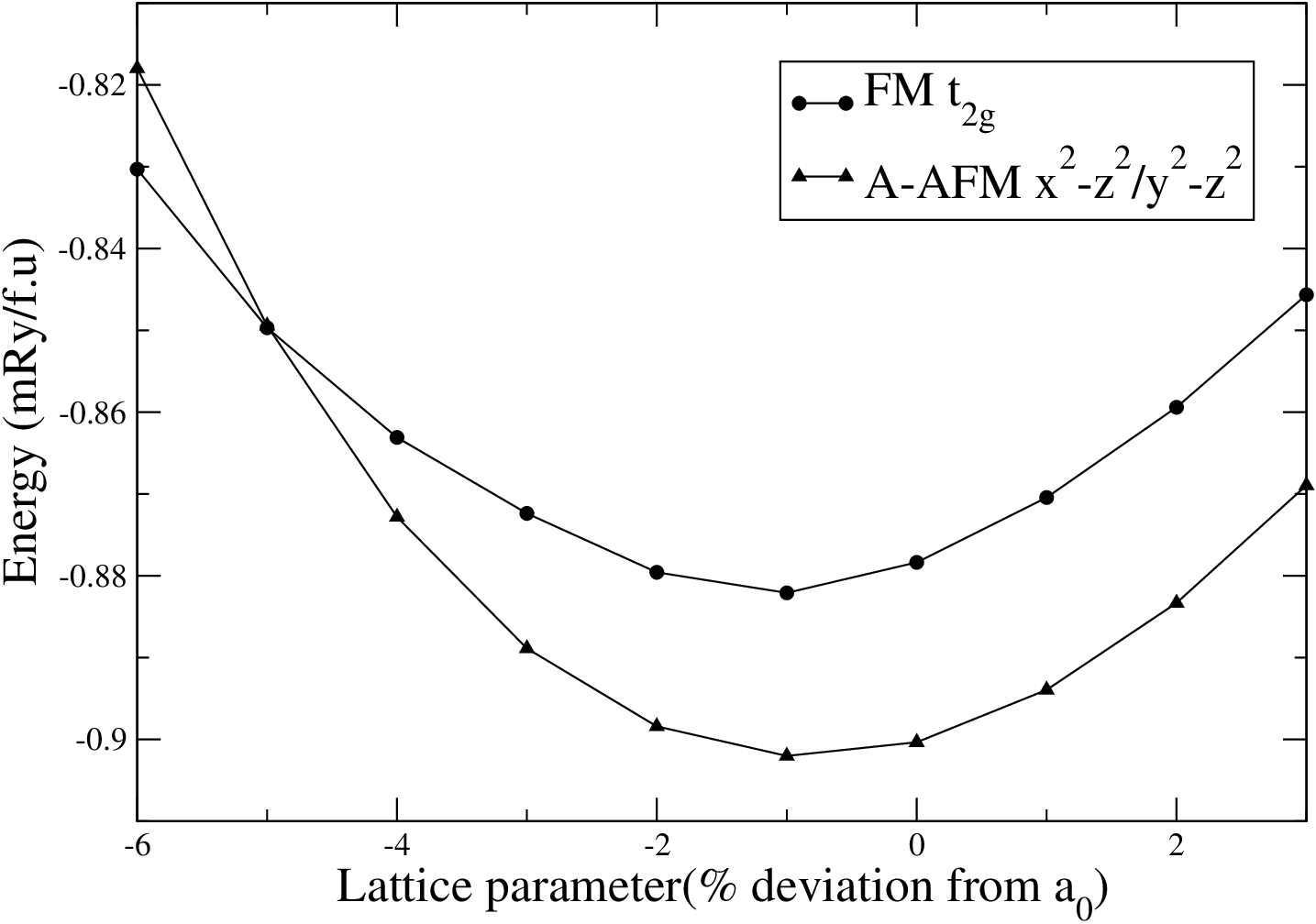}
\caption{\label{fig5} Total energies of the A-AFM $x^{2}-z^{2}$/$y^{2}-z^{2}$ 
                    orbital ordering and FM $t_{2g}$ solutions as functions of
                    the deviation of the lattice parameter $a$ from the
                    experimental one $a_{0}=7.434$a.u. We find a crossover from the $e_{g}$ localised ordered state to
                    the $e_{g}$ delocalised as $a$ is decreased.}
\end{figure}

We have also investigated the dependence of the orbital ordering on the volume
of LaMnO$_{3}$. To do so we compare total energies for different lattice
parameters relative to the experimental one. The latter is determined by
requiring that it gives the correct experimental volume of the distorted system.
We compared the energies of two scenarios: the ground state solution of the
experimental volume ($x^{2}-z^{2}$/$y^{2}-z^{2}$ orbital ordering and A-AFM
spin ordering) and the FM solution with delocalised $e_{g}$ orbits. The results
are given in Fig. \ref{fig5}. One notices that the lattice parameter
corresponding to the minimum is the same in both solutions and that it is
slightly smaller than the parameter obtained from the experimental volume of
the distorted system. Upon decreasing the volume the two curves cross a at about
-5\% of the experimental lattice parameter. Below this value the $e_{g}$
electron becomes delocalised and there is no longer orbital
ordering. The system becomes metallic too as was signalled by
the jump in the conductivity found by Loa \textit{et al.} \cite{loa}.

\section{\label{sec4} Conclusions}
We have investigated orbital ordering in cubic LaMnO$_{3}$ using the SIC-LSD
method which allows to study the localisation-delocalisation competition of
correlated electrons. Although orbital ordering in LaMnO$_{3}$ has been
ascribed to Jahn-Teller distortions of the MnO$_{6}$ octahedra we found that
this ordering can happen from purely electronic effects by spontaneous
breaking of the cubic symmetry. Once the orbital ordering sets in the 
electrostatic interaction between the O ions and the electrons on the
neighbouring Mn ions can be minimised by elongating the bonds along the lobes
of the occupied $e_{g}$ orbitals. It seems though that this coupling to the
lattice is still needed to select the correct orbital ordering giving the
observed distortions in the real LaMnO$_{3}$ system. There is therefore no
need to assume  an underlying A-AFM magnetic ordering to recover the 
orbital ordering. The latter is independent of the magnetic ordering and this
is evidenced by the much higher ordering temperature of the orbits as compared
to the spins. Although what we have found is that the lattice is
important to determine the symmetry of the ground state orbital ordering.

\end{document}